\documentclass{elsart}
\usepackage{graphics}
\usepackage{graphicx}
\usepackage{epsfig}
\usepackage{amssymb}
\usepackage[fleqn]{amsmath}

\begin{document}
\begin{frontmatter}

\title{Semiclassical investigation of revival phenomena in one
dimensional system}

\author{Zhe-xian Wang $^{1}$
\quad Eric J. Heller $^{2}$}

\address{1 Hefei National Laboratory for Physical Sciences at Microscale and Department of Physics, University of Science and
Technology of China, Hefei, Anhui 230026, P. R. China\\
2 Department of
Physics and Department of Chemistry and Chemical Biology, Harvard
University, Cambridge, Massachusetts 02138, USA}

\begin{abstract}
In a quantum revival, a localized wavepacket re-forms or "revives"
into a compact reincarnation of itself long after it has spread in
an unruly fashion over a region restricted only by the potential
energy. This is a purely quantum phenomenon, which has no classical
analog. Quantum revival, and Anderson localization, are members of a
small class of subtle interference effects resulting in a quantum
distribution radically different from the classical after long time
evolution under classically nonlinear evolution. However it is not
clear that semiclassical methods, which start with the classical
density and add interference effects, are in fact capable of
capturing the revival phenomenon. Here we investigate two different
one dimensional systems, the infinite square well and Morse
potential. In both cases, after a long time the underlying classical
manifolds are spread rather uniformly over phase space and are
correspondingly spread in coordinate space, yet the semiclassical
amplitudes are able to destructively interfere over most of
coordinate space and constructively interfere in a small region,
correctly reproducing a quantum revival. Further implications of
this ability are discussed.
\end{abstract}

\begin{keyword}
Quantum revival \sep Semiclassical \sep Infinite square well \sep
Morse potential
\PACS 03.65.Sq \sep 42.50.Md
\end{keyword}
\end{frontmatter}

\section{Introduction}
The phenomenon of "quantum revival" attracted much attention after
it was first studied in quantum electrodynamics \cite{ref01,ref02}.
The evolution of a quantum wave packet in a general smooth potential
has at least three regimes. First, an initially localized packet
will evolve following classical mechanics for a time, in the sense
that the mean position and momentum of the wave packet follow
classical laws. More than that, the spreading of the wave packet
follows an analogous classical distribution with  appropriate
initial position and momentum densities. This is the Ehrenfest
regime.

After further evolution, after the wave packet has become
delocalized, interference effects may become important, causing the
classical distribution and the quantum wave packet to have quite
different details. Semiclassical methods however are expected to be
working well. They are based solely on classical information, but
incorporate interference effects by assigning an amplitude and phase
for the multiple classical paths which connect to each final
position:
\begin{equation}
\label{universal} {\psi(x,t) = \sum_n\sqrt{P_n(x,t)}\
e^{i\,\phi_n(x,t)/\hbar}}
\end{equation}
where $P_n(x)$ is the classical probability density for the $n^{th}
$ way of reaching $x$ give the initial classical manifold and
$\phi_n(x,t)$ is the classical action along  the $n^{th}$ path
reaching $x$. The Born interpretation, namely that $\psi(x,t)$ is a
probability amplitude, dictates that the wavefunction should go as
the square root of the classical probabilities in the correspondence
limit.

After a very long period of time, many classical periods in the case
of an oscillator, the quantum wave packet will reverse its seemingly
unorganized delocalized oscillation to neatly  rebuild into its
initial form. This is the  known quantum revival, the third regime.
Quantum revival has been widely investigated in atomic
\cite{ref03,ref04,ref05} and molecular \cite{ref06,ref07,ref08} wave
packet evolution and other quantum mechanics systems
\cite{ref09,ref10,ref11,ref12,ref13}. An excellent review on wave
packet revivals is given by Robinett \cite{ref14}. Precursors to the
full revival also exist, in which other organized probability
distributions develop \cite{ref14}.  The question addressed in this
paper is: is the third, revival regime also semiclassical? May we
think of revival in semiclassical terms after all, i.e. classical
mechanics with phase interference  included?  It is a tall order for
semiclassical sums to self cancel almost everywhere the classical
density is large, with the exception of one region where the revival
is occurring.

Time dependent semiclassical methods are exact in the limit of short
time, being equivalent to the  short time limit of the quantum
propagator. Increasing time can only degrade the results.  At long
times, the number of terms in the sum, Eq.~\ref{universal} can
become very large, and in fact the number of terms grows
exponentially in chaotic systems. This in itself does not spell the
breakdown of semiclassics. In earlier work on chaotic systems,
Tomsovic et. al. \cite{ref22} showed that semiclassical amplitudes
were doing well when more than 6000 terms were needed in the sum.
Other work justified the unexpected accuracy of the semiclassical
results \cite{ref23}. Later, Kaplan \cite{Kaplan} gave an ingenious
analysis of the breakdown with time in the case of chaotic systems,
which built on earlier the analyses \cite{ref23} indicating that
classical chaos rather surprisingly {\it aided} accurate
semiclassical propagation. The implication was that even Anderson
localization was describable semiclassically, albeit with an
astronomical number of terms in the sum, Eq.~\ref{universal}.
Quantum revival in a potential well does not involve chaotic
spreading in phase space, and thus it could be more difficult to
describe correctly semiclassically, give the arguments in the above
references about the benefits of chaotic flow.

The revival phenomenon has no purely classical analog.  At best it
is a semiclassical effect, describable in the terms of
Eq.~\ref{universal}. The classical analog of a localized wave packet
will be a continuous density of trajectories in phase space, well
localized but consistent with the uncertainly principle.  In an
anharmonic oscillator, these trajectories occupy a distribution of
energies and hence frequencies. The distribution spreads and begins
to wind itself up on a spiral (see below), with many branches at a
typical position. A smooth distribution of trajectories with a range
of  velocities and positions, after spreading evenly into the
available space, will never converge again on one locale. This seems
quite contradictory to the quantum result. Semiclassical theory can
bridge the gap between classical and quantum field, and provide a
simple and intuitive way to understand the subtle issue of quantum
revival.

In this paper we study the quantum revival in both infinite square
well and Morse potential system. These two cases are quite different
in detail. The square well is locally linear, interrupted by
discontinuities which are due to reflections at the walls. The Morse
potential is more typical, arriving at its nonlinear evolution
smoothly. Semiclassical results are analytic whenever the dynamics
is "linear". Examples are the free particle, the linear ramp
potential, and the harmonic oscillator. In each case, current
positions and momenta are linear functions of initial positions and
momenta. The square well is not in fact a linear system, because of
reflections at the walls. However, locally, the classical manifolds
evolve linearly, suffering truncation  due to the reflections (see
Figure \ref{Fig. 1}). Interestingly the square well is a case with
(globally) nonlinear time evolution clearly showing revivals, yet
because of the locally linear nature of the classical dynamics the
semiclassical formula turns out to be exact. When the semiclassical
method is approximate, the delicate cancellation of amplitudes over
wide areas is in question, and we show here by example that it is
still accurate enough to give the revivals.

\section{Theory}
Time-dependent semiclassical methods face difficulties when applied
to long revival time calculations. By their very nature, revivals
cannot happen until the classical manifolds have folded over on
themselves many times, which means the dynamics is in the deeply
nonlinear regime. Although nothing keeps semiclassical methods from
working under these conditions in principle, and practice the error
can only grow with time. If one is looking at a subtle phenomenon,
such as near exact cancellation of semiclassical amplitudes over a
wide area, the small errors could be a problem.

A convenient way to implement the semiclassical method is via
cellular dynamics \cite{ref17}, which has been proven to be accurate
and efficient for longtime implementation of semiclassical
calculations.  The basic idea is to linearize the classical dynamics
in zones small enough to make the linearization classically correct.
The zones are typically much smaller than Planck's constant in area.
In the following, a brief summary of cellular dynamics is
given. In the next section we discuss the revival in both
infinite square well and Morse potential systems. Further
speculations are given in the Conclusion.

The starting point of semiclassical method is the
Van-Vleck-Gutzwiller (VVG) propagator \cite{ref18}
\begin{align}
\label{eq01} G\left({x,x_0;t} \right)\ =& \left(\frac{1}{2\pi{\rm
i}\hbar}\right)^{1/2}\sum_{j}\left| \frac {\partial ^2
S_{j}(x,x_0)}{\partial x \partial x_0} \right|
^{1/2}\exp\left[\frac{{\rm i}S_{j}(x,x_0)}{\hbar}-\frac{{\rm
i}\nu_j\pi}{2}\right] \nonumber\\
=& \left(\frac{1}{2\pi{\rm i}\hbar}\right)^{1/2}\sum_{j}\left| \frac
  {\partial x}{\partial p_0}\right|^{-1/2}\exp\left[\frac{{\rm i}S_{j}
(x,x_0)}{\hbar}-\frac{{\rm i}\nu_j\pi}{2}\right] ,
\end{align}
where action $S(x,x_0) =\int_{\rm{0}}^t {dt'} \left[ {p\left( {t'}
\right)\dot x\left( {t'} \right) - H\left( {p\left( {t'}
\right),x\left( {t'} \right)} \right)} \right] $ is the integral of
the Lagrangian along classical trajectory from $x_0$ to $x$, and
Maslov index $\nu$ counts the number of caustic points along this
trajectory. The sum over $j$ runs over all the trajectories
connecting $x_0$ to $x$, in other words, it counts in contributions
from all the stationary phase points. Cellular dynamics begins with
a transformation of the propagator by applying the speciality of
$\delta$ function:
\begin{equation}
\label{eq02} \sum {\frac{1}{{\left. {\left( {\partial x_t /\partial
p_0 } \right)} \right|_{x = x_t } }} = \int {dp_0 \delta \left( {x -
x_t \left( {x_0 ,p_0 } \right)} \right)}}.
\end{equation}
Here $x_t \left( {x_0 ,p_0 } \right) $ is the final position
originate from initial point $\left( {x_0 ,p_0 } \right)$. The VVG
propagator can now be written as
\begin{equation}
\label{eq03} G\left( {x,x_0 ;t} \right){\rm{ = }}\left(
{\frac{{\rm{1}}}{{{\rm{2}}\pi i\hbar }}} \right)^{1/2} \int {dp_0 }
\left| {\frac{{\partial x_t }}{{\partial p_0 }}} \right|_{x_0
}^{1/2} \delta \left( {x - x_t \left( {x_0 ,p_0 } \right)}
\right)\exp \left[ {\frac{{iS\left( {x_0 ,p_0 } \right)}}{\hbar } -
\frac{{i\upsilon \pi }}{2}} \right],
\end{equation}
with the change of action $S$ as a function of $(x_0 ,p_0)$. Then we
can get the semiclassical wave function
\begin{align}
\label{eq04} \psi \left( {x,t} \right) =& \int {dx_0 G\left( {x,x_0
;t}\right)\psi \left( {x_0 ,0} \right)}\nonumber\\
  =&\left({\frac{{\rm{1}}}{{{\rm{2}}\pi i\hbar }}} \right)^{1/2} \int
{dx_0 } \int {dp_0 \left| {\frac{{\partial x_t }}{{\partial p_0 }}}
\right|^{1/2} \delta \left( {x - x_t } \right)e^{iS/\hbar  - i\nu
\pi /2} \psi \left( {x_0 ,0} \right)}.
\end{align}
It would be difficult evaluate the integral directly since it is
highly oscillatory.  However, cellular dynamics handles this
difficulty by using integration techniques similar in spirit to
Filinov methods, by dividing the  region into small cells, inserting
the identities $ 1\approx \eta \sum\limits_n {\exp [ - \alpha \left(
{x - x_n } \right)^2 ]}$ within both $x$ and $p$ space.  Then we
have
\begin{align}
\label{eq05} \psi \left( {x,t} \right) = \eta \eta '\sum\limits_n
{\sum\limits_m {\int {dx_0 \int {dp_0 \left| {\frac{{\partial x_t
}}{{\partial p_0 }}} \right|^{1/2} \delta \left( {x - x_t } \right)}
} } }e^{iS/\hbar  - i\nu \pi /2}\nonumber\\
  \times e^{ - \alpha \left( {x_0  - x_n } \right)^2  - \beta
\left( {p_0  - p_m } \right)^2 }
   e^{ - \gamma \left( {x_0  - x_i } \right)^2  + ik_i \left( {x_0  -
x_i } \right)^2 } ,
\end{align}
where the initial wave function $\psi \left( {x_0 ,0} \right) = \exp
\left[ { - \gamma \left( {x_0  - x_i } \right)^2  + ik_i \left( {x_0
- x_i } \right)^2 } \right] $ is used. If both $\alpha$ and $\beta$
are taken to be sufficiently large, and  for sufficiently many
cells, we can linearize the classical dynamics around the central
trajectory for each cell running from the initial phase space point
$\left( {x_n ,p_m } \right)$, obtaining its contribution to the
propagation of initial wave function.

In some ways cellular dynamics resembles Miller's initial value
representation (IVR)\cite{miller}, but there are important
differences. The IVR is actually numerically superior, in that if
the integral is performed the result is not the "primitive
semiclassical" Van Vleck result, but rather a uniformized version
which is capable of describing some classically forbidden processes
and of smoothing out some semiclassical singularities. In contrast,
cellular dynamics is a direct  but numerically convenient
implementation of the primitive semiclassical Green's function. The
goal of the present paper is to test the efficacy of the primitive
semiclassical propagator, but implementing an IVR would be an
interesting study.

The linearization is implemented by approximating classical action
$S$ with second order Taylor expansion and final position $x\left(
{x_0 ,p_0 } \right)$ with first order \cite {ref17}, viz.
\begin{equation}
\label{eq06}
\begin{array}{rcll}
S & \approx & S_{nmt}  + \left( {p_{nmt} m_{22}  - p_m }
\right)\left( {x_0  - x_n } \right) + p_{nmt} m_{21} \left( {p_0  -
p_m } \right)\\
    & & + \frac{1}{2}m_{12} m_{22} \left( {x_0  - x_n }
\right)^2+\frac{1}{2}m_{11}m_{21} \left( {p_0-p_m} \right)^2\\
& & + m_{12}m_{21} \left( {x_0-x_n} \right) \left( {p_0-p_m}\right)\\
\multicolumn{3}{l}{x_t(x_0,p_0) \approx x_{nmt} + m_{21}(p_0-p_m)+
m_{22}(x_0-x_n)}.
\end{array}
\end{equation}
where $S_{nmt}$, $x_{nmt}$, $p_{nmt}$ are the classical action,
final position and momentum of a trajectory originate from $ \left(
{x_n ,p_m } \right)$  respectively, and
\begin{equation}
\label{eq07} M = \left( {\begin{array}{*{20}c}
    {m_{11} } & {m_{12} }  \\
    {m_{21} } & {m_{22} }  \\
\end{array}} \right) = \left( {\begin{array}{*{20}c}
    {\partial p_t /\partial p_0 } & {\partial p_t /\partial x_0 }  \\
    {\partial x_t /\partial p_0 } & {\partial x_t /\partial x_0 }  \\
\end{array}} \right)
\end{equation}
is the Jacobian matrix of the corresponding dynamical transformation
\cite {ref17}. The substitution of equation (\ref {eq06}) into (\ref
{eq05}) will simplify the quadrature into Gaussian integration
\begin{equation}
\label{eq08}
  \psi \left( {x,t} \right) = \eta \eta '\sum\limits_n
{\sum\limits_m {\int {dx_0 \left| {\frac{{\partial x_t }}{{\partial
p_0 }}} \right|^{ - 1/2} e^{ - a\left( {x_0  - x_n } \right)^2  +
b\left( {x_0  - x_n } \right) + c} } } } ,
\end{equation}
with the coefficients
\begin{align}
\label{eq09}
  a = & \alpha  + \gamma  + \beta \left( {\frac{{m_{22}
}}{{m_{21} }}} \right)^2  - \frac{i}{\hbar }\left(
{\frac{1}{2}\frac{{m_{11}
m_{22}^2 }}{{m_{21} }} - \frac{1}{2}m_{12} m_{22} } \right),\nonumber\\
b = & \frac{{2\beta m_{22} }}{{m_{21}^2 }}\left( {x - x_{nmt} }
\right) - 2\gamma \left( {x_n  - x_i } \right) + ik_i \nonumber\\
&+ \frac{i}{\hbar }[\left( {m_{12}  - \frac{{m_{11} m_{22}
}}{{m_{21} }}} \right)\left( {x - x_{nmt} } \right) - p_m ],\nonumber\\
c =&  - \frac{\beta }{{m_{21}^2 }}\left( {x - x_{nmt} } \right)^2  -
\gamma \left( {x_n  - x_i } \right)^2 - \frac{{i\nu \pi }}{2} + ik_i
\left( {x_n  - x_i } \right)\nonumber\\
& + \frac{i}{\hbar }[S_{nmt}  + p_{nmt} \left( {x - x_{nmt} }
\right) + \frac{{m_{11} }}{{2m_{21} }}\left( {x - x_{nmt} }
\right)^2 ].
\end{align}
The equation (\ref {eq05}) can now be analytically evaluated
\begin{equation}
\label{eq10} \psi \left( {x,t} \right) = \eta \eta '\sum\limits_n
{\sum\limits_m {\sqrt {\frac{\pi }{{am_{21} }}} e^{b^2 /4a + c}}},
\end{equation}
and it is easy to implement.

\section{Results and discussions}
\begin{figure}[htbp]
\begin{center}
\includegraphics[width=1.0\textwidth,bb=24 26 474 236]{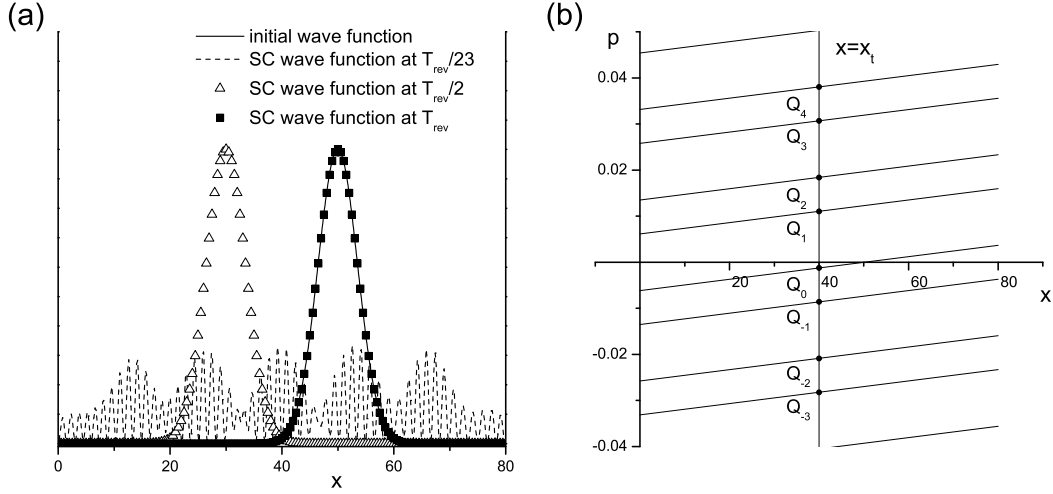}
\caption{(a) Semiclassical wave functions evolve in infinite square
well at different times. (b) Partial classical manifold of an
initial $\delta$ function evolves in infinite square well at time
$T_{rev}$.} \label{Fig. 1}
\end{center}
\end{figure}

\begin{figure}[htbp]
\begin{center}
\includegraphics[width=1.0\textwidth,bb=30 34 517 224]{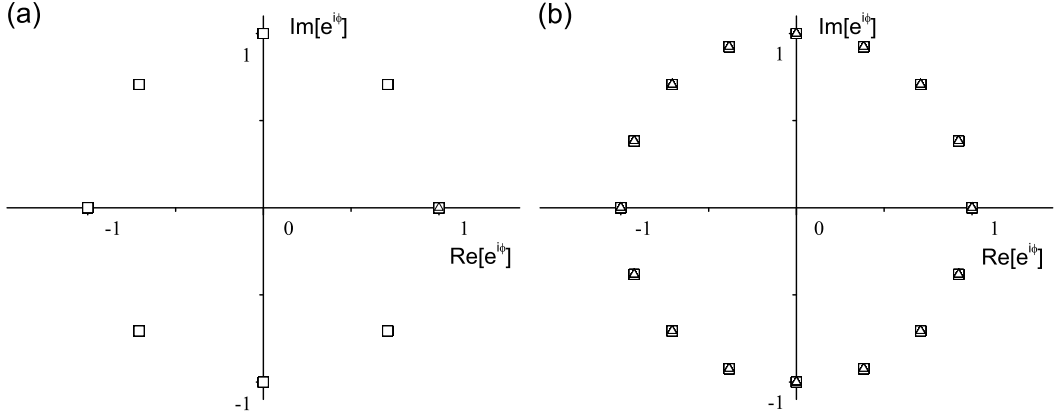}
\caption{The distribution of exponential function in complex plane
for position (a)$x=50$; (b)$x=11$. The square and triangle indicate
phase terms come from the first and second exponential function in
equation (\ref {eq17}) respectively.} \label{Fig. 2}
\end{center}
\end{figure}

\begin{figure}[htbp]
\begin{center}
\includegraphics[width=0.8\textwidth,bb=18 18 301 220]{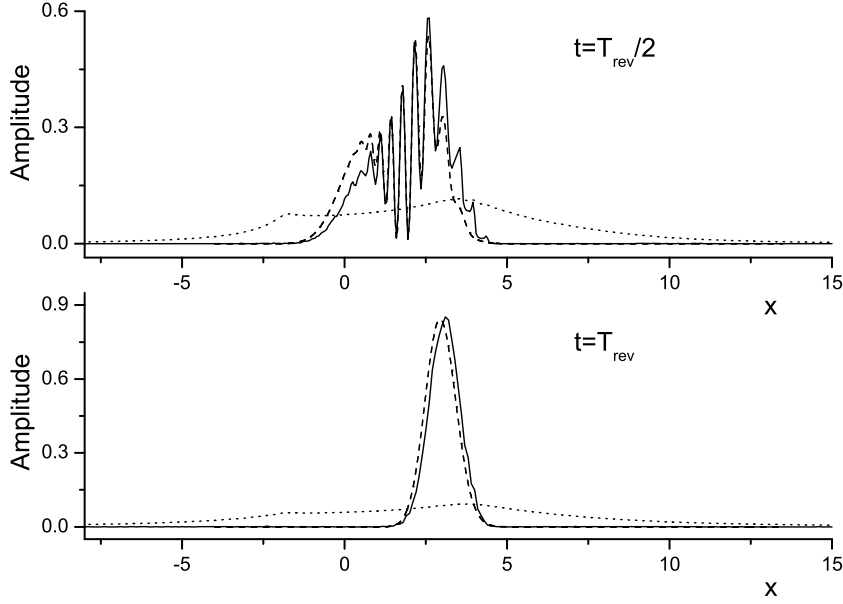}
\caption{Wave functions and classical distribution probabilities in
Morse potential at time $T_{rev} /2$ and $T_{rev}$. All the
functions plot in this figure are normalized. Solid line:
Semiclassical wave functions; Dash line: Exact FFT wave functions
calculated by Split-Operator method \cite {ref19};Dot line:
Classical  density in coordinate space  which evolves from the  initial density. } \label{Fig. 3}
\end{center}
\end{figure}

\begin{figure}[htbp]
\begin{center}
\includegraphics[width=1.0\textwidth,bb=0 0 1750 600]{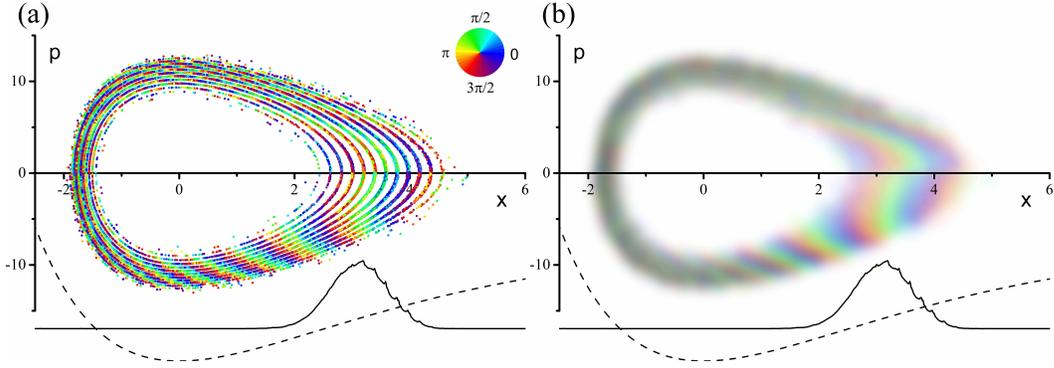}
\caption{(a)Phase space diagram for Wigner transformed Gaussian
evolves in Morse potential at time $T_{rev}$. The color indicates
different value of phase (include classical action $S$ and Maslov
phase) divided by $2\pi$. (b) The blurred version of figure (a).
Solid line: semiclassical wave function at time $T_{rev}$; Dash
line: Morse potential.} \label{Fig. 4}
\end{center}
\end{figure}

\begin{figure}[htbp]
\begin{center}
\includegraphics[width=1.0\textwidth,bb=25 23 434 213]{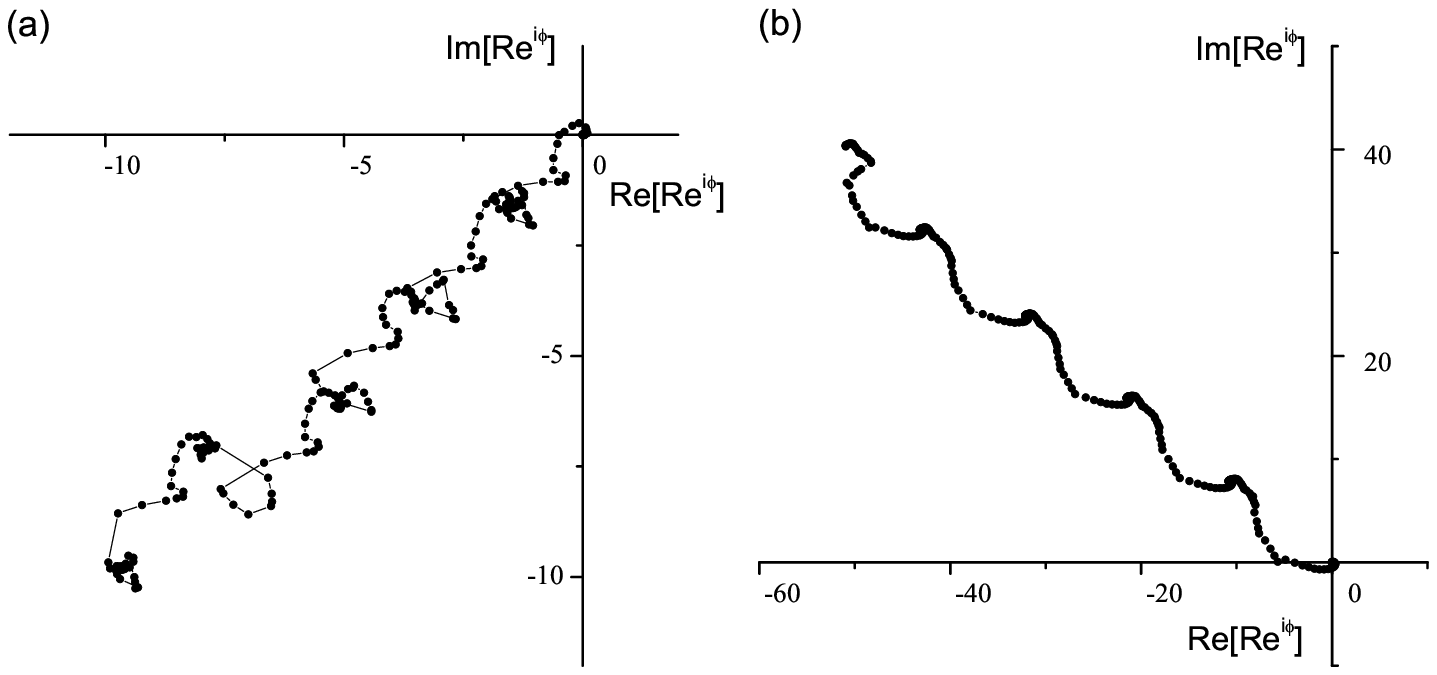}
\caption{(a)Vector chain for $x=-1.6$. (b)Vector chain for $x=3.1$}
\label{Fig. 5}
\end{center}
\end{figure}
In this section we will analyze the quantum revival in the infinite
square well and Morse potential in detail. First we look at the
infinite square well system, which has been well studied at many
levels and from many points of view \cite {ref11,ref12,ref13}. We
take the initial Gaussian of the form
\begin{equation}
\label{eq11}
\psi \left( {x_0 ,0} \right) = \sqrt {\gamma /\pi }
\exp \left[ { - \gamma \left( {x_0  - x_i } \right)^2  + ik_i \left(
{x_0 - x_i } \right)^2 } \right]
\end{equation}
and the system Hamiltonian is
\begin{equation}
\label{eq12}
 H = p^2 /2m + V\left( x \right),{\rm{  }}V\left( x
\right) = \left\{ {\begin{array}{*{20}c}
   {0,{\rm{       }}0 < x < L}  \\
   {\infty ,{\rm{   }}x \le 0,x \ge L}  \\
\end{array}} \right..
\end{equation}
In infinite square well system, by expanding the evolving wave
function with eigen states, the revival time $T_{rev}  = 4mL^2
/\hbar \pi $ can be analytically determined \cite {ref12}, and it
only depends on electron mass and the width of the well. In all the
calculations $ m = 1,{\rm{ }}\hbar  = 1$ are used for simplicity.
With parameters $\gamma  = 0.02,{\rm{ }}k_i  = 2,{\rm{ }}x_i  = 50 $
and $L = 80 $ we compute the semiclassical wave function at
different times using cellular dynamics, we take 100 cells equally
spaced in $x$ from 20 to 80 and 1500 cells in $p$ from 0.8 to 3.2.
One should pay attention to the Maslov phase here, in hard wall
limit it is a multiple of $\pi$ instead of $\pi/2$. As shown in
Fig~\ref{Fig. 1} (a), the wave packet quickly spread over the well
after first several classical periods, and at time $t = T_{rev} /2$
the wave function is a mirror image of initial wave function, then
after revival time $T_{rev}$ the wave function is perfectly rebuilt
into initial wave packet.  The reason we suggest for this
astonishing relocalize of wave packet is the interference between
different contributing classical trajectories. In the following we
unfold our discussions.

As the revival in infinite square well is independent of the shape
of the wave packet, we can take a quite narrow initial wave
function, such as $ \psi \left( {x_0 ,0} \right) = \delta \left(
{x_0  - x_i } \right)$, sits at $x_i  = 50$, of course it will
rebuild itself at time $t = T_{rev}$. Then from equation (\ref
{eq04}) we see the wave function directly connect to the
semiclassical propagator,
\begin{align}
\label{eq13} \psi \left( {x,t} \right) = &\int {dx_0 G\left( {x,x_0
;t} \right)\psi \left( {x_0 ,t} \right)}\nonumber\\
& = \int {dx_0 G\left( {x,x_0 ;t} \right)\delta \left( {x_0  - x_i }
\right)}  = G\left( {x,x_0 ;t} \right).
\end{align}
Referring to the existing works on Feynman path integral in infinite
square well \cite {QM,Handbook}, the semiclassical propagator can be
written as a summation of contributions from all the stationary
phase points
\begin{align}
\label{eq14}G\left( {x,x_0 ;t} \right)=&\sqrt {{m \over {2\pi i\hbar
t}}} \left[ {\sum\limits_{n =  - \infty }^\infty  {\exp \left(
{{{im\left( { - x - x_0  + 2nL} \right)^2 } \over {2\hbar t}} -
i\left| {2n - 1}
\right|\pi } \right)} }   \right. \nonumber\\
&\left. + {\sum\limits_{n =  - \infty }^\infty  {\exp \left(
{{{im\left( {x - x_0  + 2nL} \right)^2 } \over {2\hbar t}} - i\left|
{2n} \right|\pi } \right)} }  \right] \nonumber\\
=&\sqrt {{m \over {2\pi i\hbar t}}} \left[{\sum\limits_{n =  -
\infty }^\infty  {\exp \left( {{{im\left( {x - x_0  + 2nL} \right)^2
} \over {2\hbar t}}} \right)} }\right. \nonumber\\
&\left. -{\sum\limits_{n =  - \infty }^\infty  {\exp \left(
{{{im\left( { - x - x_0  + 2nL} \right)^2 } \over {2\hbar t}}}
\right)} }\right]
\end{align}
Fig~\ref{Fig. 1} (b) shows part of the manifold at time $T_{rev}$
which evolves from initial $\delta$  wave function, the intersection
of $x = x_t$ with manifold produces stationary phase points. The two
sums in Eq.~\ref{eq14} correspond to stationary phase points with
classical trajectories bouncing off the wall by even and odd times
respectively. To simplify the Eq.~\ref{eq14}, we use the Jacobi
theta function $ \vartheta _3 \left( {z,T} \right) = \sum\limits_{n
= - \infty }^\infty  {\exp \left[ {i\left( {\pi n^2 T + 2nz}
\right)} \right]} $ and its important property \cite {Mathbook}
\begin{equation}
 \label{eq15}\vartheta _3 \left(
{z,T} \right) = \sqrt {i/T} \exp \left( {z^2 /i\pi T}
\right)\vartheta _3 \left( {z/T, - 1/T} \right),
\end{equation}
the semiclassical propagator becomes
\begin{align}
\label{eq16} G\left( {x,x_0 ;t} \right)=&{1 \over {2L}}\left[
{\vartheta _3 \left( {{{\pi \left( {x - x_0 } \right)} \over
{2L}},{{ - \pi \hbar t} \over {2mL^2 }}} \right) - \vartheta _3
\left( {{{ - \pi \left( {x + x_0 } \right)} \over {2L}},{{ - \pi
\hbar t} \over {2mL^2 }}} \right)} \right]\nonumber\\
=& {1 \over {2L}}\sum\limits_{n =  - \infty }^\infty  {\exp \left(
{{{ - in^2 \pi ^2 \hbar t} \over {2mL^2 }}} \right)} \left[ {\exp
\left( {{{in\pi \left( {x - x_0 } \right)} \over L}} \right)}
\right.\nonumber\\
&\left.- {\exp \left( {{{ - in\pi \left( {x + x_0 } \right)} \over
L}} \right)} \right]\nonumber\\
=&{2 \over L}\sum\limits_{n = 1}^\infty  {\exp \left( {{{ - in^2 \pi
^2 \hbar t} \over {2mL^2 }}} \right)} \sin \left( {{{n\pi x_0 }
\over L}} \right)\sin \left( {{{n\pi x} \over L}} \right).
\end{align}
This is identical to the  usual quantum propagator in infinite square
well. At the revival time $T_{rev}  = 4mL^2 /\hbar \pi$ the wave
function can be rewritten as
\begin{align}
\label{eq17} \psi \left( {x,T_{rev} } \right) =& G\left( {x,x_0
;T_{rev} } \right)\nonumber\\
=& {1 \over {2L}}\sum\limits_{n =  - \infty }^\infty  {e^{ - i2n^2
\pi } } \left[ \exp \left( {{{in\pi \left( {x - x_0 } \right)} \over
L}} \right)\right. \nonumber\\
& \left. + \exp \left( {{{ - in\pi \left( {x + x_0 } \right)} \over
L} + i\pi } \right) \right]\nonumber\\
=& {1 \over {2L}}\sum\limits_j {e^{i\phi _j } } .
\end{align}
We can compare analytically the difference between low and high
amplitude points of wave function. Taking $x_1=20,x_2=50$ for
example, we find that for the high amplitude position at $x=50$, the
exponential functions $ \exp \left( {i\phi _j } \right)$ in the
summation distribute  uniformly in complex plane. There are only
eight phase terms [see Fig~\ref{Fig. 2} (a)] in the sum. We need to
distinguish the phase terms come from different exponential function
in Eq.~\ref{eq17}. The second exponential function gives out all 8
different terms distribute symmetrically around the circle so that
they will cancel each other, whereas, the first exponential function
only gives out $\phi=0$ terms, they will build up big contributions
and give out high amplitude. For the low amplitude point $ x=20$,
however, both exponential functions generate 16 symmetrically
distributed terms [see Fig~\ref{Fig. 2} (b)] on the unit circle and
therefore the summation approaches zero. Hence the interference
between part of different classical trajectories yields the revival
of wave packet.

Now we come to see a more general system, the Morse potential. It is
also a widely used model in many fields. We take $V\left( x \right)
= D\left[ {1 - \exp \left( {-\lambda x} \right)} \right]^2$ with $D
= 150,\lambda = 0.288$, its revival time $T_{rev}=2m\pi /\left(
{\hbar \lambda } \right)^2$ can be derived by expanding the wave
function with eigen functions of Morse potential, too [see Appendix
A]. With 300 cells in $x$ and 600 cells in $p$ been used in the
calculation, the semiclassical wave functions originate from $\psi
\left({x_0 ,0} \right)=\sqrt {\gamma /\pi } \exp \left[{-\gamma
\left({x_0-x_i} \right)^2} \right]$ $ \left({\gamma=2,x_i=3.5}
\right)$ are pictured in Fig~\ref{Fig. 3}. Comparing to the FFT
exact wave functions we can see semiclassical wave functions agree
well for different time scales. In Fig~\ref{Fig. 3} we plot the
normalized classical coordinate space density arising from the
initial classical distribution. Since the semiclassical result
consists of the square root of classical probabilities multiplied by
phase terms and added together, but it is easy to construct its
purely classical result by removing the phase terms, and squaring
and adding all the square root classical densities.

It is surprising that despite the fact that the classical
trajectories are spread all over the available phase space and
coordinate space, the semiclassical approximation can still build a
localized wave packet at the revival time $T_{rev}$.

In order to demonstrate the relationship between semiclassical wave
function and classical information carried by trajectories, we first
Wigner transform the initial Gaussian distribution $\psi \left(x
\right)=\sqrt{\gamma /\pi} \exp \left[{-\gamma \left(x-x_i \right)^2
}\right]$,
\begin{align}
 W\left({x,p} \right)= &\frac{1}{{\pi \hbar}}\int_{-\infty }^\infty {\psi ^* \left({x - s} \right)\psi \left({x + s} \right)e^{i2ps/\hbar} ds} \nonumber\\
=&\frac{\gamma}{{\pi ^2 \hbar}}\int_{-\infty}^\infty {e^{-\gamma \left({x-x_i-s} \right)^2 } e^{-\gamma \left({x-x_i+ s} \right)^2} e^{i2ps/\hbar} ds} \nonumber\\
=&\frac{\gamma}{{\pi ^2 \hbar}}\int_{-\infty }^\infty {e^{-2\gamma \left({x-x_i} \right)^2-2\gamma s^2+ i2ps/\hbar} ds} \nonumber\\
=&\sqrt {\frac{\gamma }{{2\pi ^3 \hbar ^2 }}} e^{ - p^2 /2\gamma\hbar- 2\gamma \left( {x - x_i } \right)^2 } , \nonumber\\
\end{align}
which remains a Gaussian. Then we plot the evolution of this Wigner
distribution in phase space after revival time $T_{rev}$ in
Fig~\ref{Fig. 4}. The starting swarm of classical trajectories
emerges as an elliptical disk; as time evolves this ellipse
stretches and twists, forming a large whorl. (Indeed, the time
evolution of the phase space is that of an area preserving twist
map).

It might appear that the vertical sections of the classical
manifolds on the left and right sides of the whorl would dominate
the contribution to the semiclassical wave function for two reasons:
First, the prefactor $1/\sqrt {\left| {\partial x/\partial p_0 }
\right|}$ in VVG propagator in equation (\ref {eq01}) is large for
this part of the manifold. This is because the density of the
distribution is proportional to the probability $1/\sqrt{\left|
{\partial x/\partial p_0} \right|}$ of classical particles locating
at those regimes. Second, these particles have similar classical
actions [see Fig~\ref{Fig. 4} (a)]. In Appendix B we prove the
classical action difference between two points equals the enclosed
area of the manifold. Near the fold regimes small enclosure areas
lead to similar classical actions. The abrupt changes of color at
the turning points indicate the change of Maslov index at those
points. The combination of these two factors yields large result
refers to equation (\ref {eq01}). In Fig~\ref{Fig. 4} (b) we blurred
the phase space diagram for an intuitive view. The whorl average out
and give neutral gray colors everywhere except where the revival is
occurring. When combined along vertical lines, those regions with
monochromatic bright colors will give out the revival wave packet.

Nevertheless, one could still doubt why we don't get a high
amplitude wave function at positions of folds on left side, they
also meet the conditions list above. To compare the difference, we
write the formula of the wave function into a compact form:
\begin{align}
 \psi \left( {x,t} \right) = &\int {dx_0 G\left( {x,x_0 ;t} \right)\psi \left( {x_0 ,0} \right)}  \nonumber\\
= &\left( {\frac{1}{{2\pi i\hbar }}} \right)^{1/2} \int {dx_0 \sum\limits_j {\left| {\frac{{\partial x}}{{\partial p_0 }}} \right|^{ - 1/2} \exp \left[ {\frac{{iS_j \left( {x,x_i } \right)}}{\hbar } - i\upsilon _j \pi } \right]} } \psi \left( {x_0 ,0} \right) \nonumber\\
= &\int {dx_0 Re^{i\phi } }  = \sum\limits_n {R_n e^{i\phi _n } \Delta x_0 } . \nonumber\\
\end{align}
We can approximate the quadrature numerically by a finite sum
of complex vector. We divide $x$ space into hundreds of sections,
and evaluate the vector separately in each section. By drawing each
vector from the tips of previous vector, the summation will form a
chain, and the line drawn from first point to the end point
represent the quadrature. We draw two chains respectively for
$x=-1.6$ and $x=3.1$ in Fig~\ref{Fig. 5}.

For the low amplitude region $x=-1.6$, in the vicinity of
destructive interference, the chain circles continuously, and
results in a small total vector [see Fig~\ref{Fig. 5} (a)]. This
indicates the phase of stationary phase points changes only slightly
and continuously, leading to destructive interference between
classical trajectories and a small amplitude of wave function. A
different situation applies in Fig~\ref{Fig. 5} (b) for the position
$x=3.1$. Here the small phase difference between stationary points
accumulates a persistent growth of the total vector, viz. the
constructive interference produces high amplitude of wave function.

\section{Conclusions and discussion}
Whenever and wherever they apply, semiclassical methods can be
extremely useful not only in computations, but in providing an
underlying intuition for quantum phenomena. Here we have shown that
something so subtle as a quantum revival still has classical
underpinnings, as seen by the successful construction of the
phenomenon using only classical mechanics as input. Semiclassical
methods are accurate enough to describe the quantum revival
phenomenon. The quantum revival phenomenon does not stem from an
accumulation of classical trajectories. Rather, the classical
trajectories are rather uniformly spread, and it is through
destructive interference of the semiclassical amplitudes that the
wave function is canceled in most places.

Of course, this is a momentary phenomenon, in the sense that beyond
the revival time the way packet will again begin to spread and
become quite delocalized quantum mechanically. However, imagine the
following scenario: at the moment of a revival, with the wave packet
built up on one side of the potential, suppose a time-dependent
barrier is erected, preventing the wave packet from any immediate
penetration beyond the barrier. If the barrier remained up, and the
potential were sufficiently asymmetric and designed properly the
quantum mechanical wave packet would remain on one side of the
barrier forever. This raises an even more interesting and
challenging question: When the time-dependent barrier was raised,
this traps classical manifolds on the ``empty" side of the barrier.
Presumably at the moment of the trapping, the semiclassical wave
function would indeed be correctly exceedingly small at that point,
but for how long could this semiclassical result correctly describe
the fact that the wave function never reappeared in that region? One
could call this the ``semiclassical propagation of nothing". That is
to say, abstracting this a little further, suppose you begin with a
quite complex set of classical manifolds, interpreted
semiclassically, which gives essentially zero semiclassical wave
function everywhere. Now, the continued semiclassical propagation of
these manifolds should continue to give a vanishingly small
function. Any errors in the semiclassical propagation will cause
wave function amplitude to appear incorrectly.

While we cannot fully explain the situation here, we believe this
phenomenon may be affecting the quantum classical correspondence in
branched electron flow \cite {ref20}. Branched electron flows are
usually ascribed to a purely classical effect \cite {ref20,ref21};
however, classical and quantum electron flows begin to disagree,
with some branches suddenly missing in the quantum result as
compared to the classical, as one moves further and further from the
source of electrons. In the future we hope to verify our conjecture
that the missing branches are an effect of destructive interference
of classical trajectories, by using semiclassical methods.

\section{Acknowledgements}
One of us (Z. X. Wang) would like to acknowledge helpful discussions
with Brian Landry. This work was supported in part by the National
Natural Science Foundation of China (Grant Nos.10574121 and
10874160), '111' Project, Chinese Education Ministry and Chinese
Academy of Sciences.
\appendix
\section{Appendix A}
\label{Appendix A}
 For Morse potential $V\left( x \right) = D\left[{1-\exp \left({-\lambda x} \right)}
\right]^2$, one can express the time dependent wave function in
terms of eigen functions $ \varphi _n \left(x \right)$ , via
\begin{equation}
\label{eqA1}
 \psi \left( {x,t} \right) = \sum\limits_{n = 0}^\infty  {a_n
\varphi _n \left( x \right)e^{ - iE_n t/\hbar } } ,
\end{equation}
where the eigen values are $E_n  = \alpha \left( {n + 1/2} \right) -
\beta \left( {n + 1/2} \right)^2$ with $\alpha  = \hbar \lambda
\sqrt {2D/m}$, $\beta  = \hbar ^2 \lambda ^2 /2m$. The revival
condition $\psi \left( {x,T} \right) = \psi \left( {x,0} \right)$
 requires
\begin{equation}
\label{eqA2}
 E_n T = \left[ {\alpha \left( {n + 1/2} \right) - \beta
\left( {n + 1/2} \right)^2 } \right]T = 2M_n \pi ,
\end{equation}
where $M_n$ are integers. Make a subtraction of adjacent $n$ of
equation (\ref {eqA2}) gives
\begin{equation}
\label{eqA3} \left( {\alpha  - 2\beta n - 2\beta } \right)T = 2K_n
\pi ,
\end{equation}
with $K_n$ are also integers. Then apply the subtraction of adjacent
of equation (\ref {eqA3}) again, we get the equation for the
shortest revival time $T_{rev}$ is
\begin{equation}
\label{eqA4} 2\beta T_{rev}  = 2\pi .
\end{equation}
So we have the revival time $T_{rev}  = \pi /\beta  = 2m\pi /\left(
{\hbar \lambda } \right)^2$.

 \section{Appendix B}
\begin{figure}[htbp]
\begin{center}
\includegraphics[width=0.5\textwidth,bb=26 16 240 189]{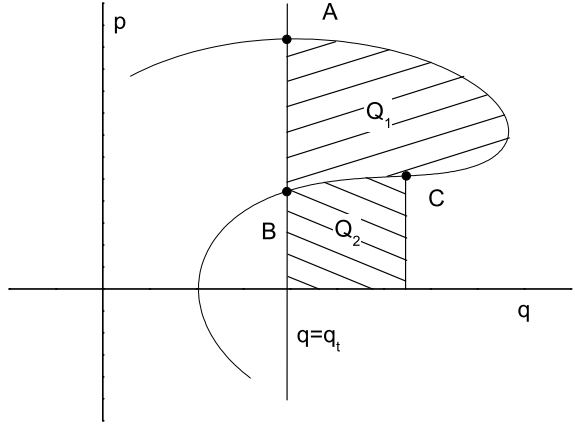}
\caption{Classical manifold. The area contained between
intersections of the manifold $p(q)$ and a position state (vertical
line $q=q_t$) is $Q_1$.} \label{Fig. B1}
\end{center}
\end{figure}

\label{Appendix B} We ought to prove the difference of classical
action $S_A$ and $S_B$ equals to the shade area $Q_1$. First we look
at point $B$ and $C$. From classical action formula $S = \int
{p\left( q \right)dq}  + \int {H\left( {p,q} \right)dt}$ we have $
\partial S/\partial q = p$, thus the action difference from $B$ to
$C$ is
\begin{equation}
\label{eqB1}
 S_C  - S_B  = \int_B^C {\frac{{\partial S}}{{\partial
q}}dq} = \int_B^C {pdq}  = area{\rm{ }}Q_2 .
\end{equation}
Then, in a similar way
\begin{equation}
\label{eqB2}
 S_A  - S_B  = \int_B^A {\frac{{\partial S}}{{\partial
q}}dq} = \int_B^A {pdq}  = area{\rm{ }}Q_1 .
\end{equation}

\appendix
 \label{}

\end{document}